\documentclass{ws-mpla}

\begin{document}

\newcommand \Pomeron {I\!\!P}

\markboth{L. Frankfurt, V. Guzey, M. Strikman}
{Nuclear shadowing and extraction of $F_2^p-F_2^n$  at small $x$ from 
electron-deuteron collider data}

\catchline{}{}{}{}{}

\title{NUCLEAR SHADOWING IN INCLUSIVE AND TAGGED DEUTERON STRUCTURE 
FUNCTIONS AND EXTRACTION OF $F_2^p-F_2^n$ AT SMALL $x$ FROM ELECTRON-DEUTERON
COLLIDER DATA}

\author{\footnotesize L. FRANKFURT}
     
\address{School of Physics and Astronomy, Tel Aviv University, \\
69978, Tel Aviv, Israel\\
frankfur@lev.tau.ac.il}

\author{V. GUZEY}

\address{Institut f{\"u}r Theoretische Physik II, Ruhr-Universit{\"a}t Bochum,\\
D-44780 Bochum, Germany\\
vadimg@tp2.rub.de}

\author{M. STRIKMAN}
\address{Department of Physics, the Pennsylvania State University, \\
State College, Pennsylvania 16802, USA\\
strikman@phys.psu.edu}

\maketitle

\pub{Received (Day Month Year)}{Revised (Day Month Year)}

\begin{abstract}

We review predictions of the theory of leading 
twist nuclear shadowing for inclusive unpolarized and polarized
deuteron structure functions $F_2^D$, $g_1^D$ and $b_1^D$ and 
for the tagged deuteron structure function $F_2^D(x,Q^2,\vec{p})$.
We analyze the possibility to extract the neutron structure function
$F_2^n$ from electron-deuteron data and demonstrate that an account of
leading twist nuclear shadowing leads to large corrections for  the
extraction of $F_2^n$ from the future deuteron collider data both in 
the inclusive and 
in the tagged structure function modes. We suggest several strategies to
address  the extraction of $F_2^n$ and to measure at the same time
the effect of nuclear shadowing via the measurement of the distortion of 
the proton spectator spectrum in the semi-inclusive $e D \to e^{\prime}NX$ process.
We address the issue of the final state interactions in the $e D \to e^{\prime}NX$ process
and examine how they affect the extraction of $F_2^n$.

\end{abstract}

\ccode{PACS Nos.:16.60.Hb, 24.85.+p, 25.30.Dh}

\section{Introduction}
\label{sec:intro}

The nuclear shadowing phenomenon  is one of key elements  in
the modern understanding of the geometry of hadron-nucleus and 
nucleus-nucleus collisions. A reliable theory of nuclear shadowing 
in hard processes is important for  establishing the
universality of hard QCD processes.  

The initial and rather successful theory of this phenomenon
was based on the optical (Glauber) approximation.\cite{Glauber:1955qq}
However, at high energies, the picture of consecutive collisions becomes inapplicable 
as a consequence of the increase of  distances involved
in high energy processes with increasing 
energies.\cite{Feinberg,Gribov:1965hf,Ioffe:1969kf}

In addition, the eikonal (optical) approximation to
inelastic processes violates energy-momentum conservation.
In the late 60's, V.~Gribov developed an approach to nuclear shadowing,
which accounts for the specifics of high energy
processes.\cite{Gribov}
This approach leads to the expression for the total hadron-nucleus
cross section which is close to the Glauber approximation,
but with an  additional effect -- inelastic shadowing. 
For the deuteron target, nuclear shadowing is unambiguously related
to the cross section of diffraction in hadron-nucleon collisions.\cite{Gribov}

The investigation of deep inelastic processes off nuclei is another challenge.
In the case of scattering off the deuteron, a generalization of Gribov's
ideas gave a possibility to evaluate nuclear shadowing in 
terms of diffractive parton densities of the nucleon.\cite{Frankfurt:1998ym}

One of applications of the theory of nuclear shadowing is the possibility to
extract the unpolarized $F_2^n$ and polarized $g_1^n$  neutron
 structure functions from electron-deuteron data.

The measurement of the non-singlet difference of the proton and neutron structure functions,
$F_2^p-F_2^n$, by using the deuteron beam and spectator tagging, is one of the important
 components of the planned physics program of the
electron-ion collider (EIC).\cite{EIC} The main goal of the measurement is to study the 
flavor dependence of parton distribution functions (PDFs) in a wide kinematic region, 
including the low-$x$ region.
In particular, the collider kinematics will allow to probe the values of Bjorken $x$, 
$x \approx 5 \times 10^{-4}$ for $Q^2 \geq 1$ GeV, which are a factor of ten smaller than in
the available fixed-target data.\cite{Arneodo:1996qe}
In addition, the measurement of $F_2^p-F_2^n$ is a unique way to investigate
nuclear shadowing of non-singlet (valence) PDFs in the non-vacuum channel.
Since at small $x$ the $F_2^p-F_2^n$ difference is compatible to the predicted nuclear shadowing
correction to the deuteron structure function $F_2^D$ 
(a few percent effect),\cite{Badelek:1991qa,Nikolaev:1990yw,Melnitchouk:1992eu,Melnitchouk:1993vc,Melnitchouk:1995am,Piller:1995kh} 
the correct extraction of $F_2^p-F_2^n$ from deuteron data requires an account for nuclear shadowing.
We demonstrate the usefulness of the investigation of the deuteron tagged structure
functions  for the extraction of the  nonsinglet flavor distributions.

The accurate measurement of nuclear shadowing in nuclear PDFs and understanding of its origins 
are of key importance for establishing the geometry of heavy-ion physics.
This impacts the present and forthcoming RHIC data and
especially the future LHC experiments. 
Therefore, it is crucial to determine the value of 
the nuclear shadowing correction, as well as its uncertainties, 
both to the inclusive structure function $F_2^{D}$ and to the tagged structure function,
when the spectator proton 
(a proton with momentum $\le 0.1$ GeV/c  in the deuteron rest frame) 
 is detected ensuring the kinematics 
maximally close to the scattering off a free nucleon.

This paper is organized as follows.
In Sec.~\ref{sec:f2d}, we discuss our predictions for the nuclear shadowing correction
to the unpolarized $F_2^D$ and polarized $g_1^D$ deuteron structure functions. The roles of
nuclear shadowing and final state interactions in the tagged unpolarized 
deuteron structure function are discussed in Sec.~\ref{sec:tagged}. We summarize and
discuss our results in Sec.~\ref{sec:conclusions}.

\section{Leading twist shadowing and inclusive deuteron structure functions}
\label{sec:f2d}

In part, this review is based on Ref.~15.
For the comprehensive review of hard processes with the deuteron, we 
refer the reader to Ref.~16.
The leading twist theory of nuclear shadowing was developed in Ref.~6
and later
applied to deep inelastic scattering (DIS) on nuclear targets 
in Refs.~17-20.
The theory of nuclear shadowing of nuclear PDFs and structure functions\cite{Frankfurt:1998ym} 
is based on Gribov's  connection between 
the nuclear shadowing correction to the total hadron-deuteron
 cross section and the cross section of
diffraction off a free nucleon,\cite{Gribov}
 Collins's factorization theorem for hard diffraction in 
DIS\cite{Collins:1997sr} and QCD analyses of HERA data on hard inclusive 
diffraction.\cite{ZEUS:1994,H1:1994}

On the qualitative level, Gribov's connection of nuclear shadowing to diffraction 
can be understood as a consequence of the interference between the amplitudes for 
diffractive scattering of the projectile  off
the proton and off the neutron of the deuterium target.
Such interference is possible for small $x$, $x \leq 5 \times 10^{-2}$,
when the minimum momentum transfer to the nucleon, $\sim x m_N$, is smaller than the 
average nucleon momentum in the deuteron.

\begin{figure}[h]
\begin{center}
\epsfig{file=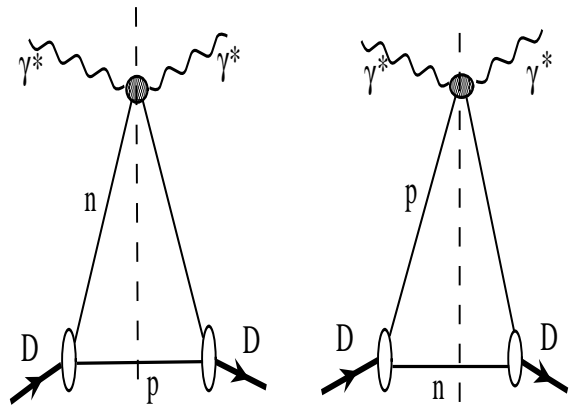,width=9cm,height=7cm}
\vskip 0cm
\caption{The impulse approximation to the deuteron structure function $F_2^D$.}
\label{fig:ia}
\end{center}
\end{figure}

\begin{figure}[h]
\begin{center}
\epsfig{file=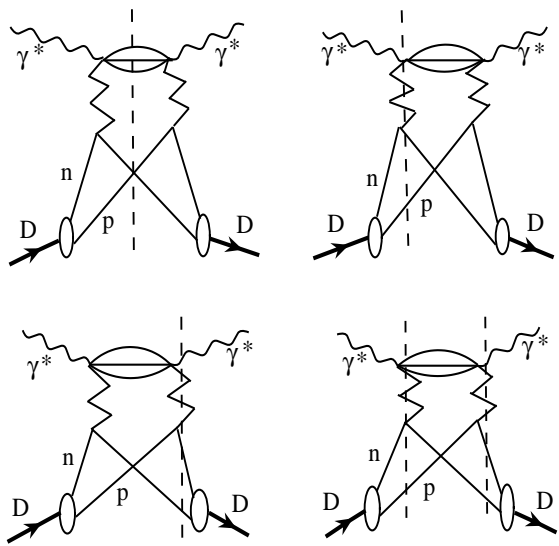,width=9cm,height=8cm}
\vskip 0cm
\caption{All possible unitary cuts of the interference diagram giving rise to the nuclear shadowing 
correction to $F_2^D$.}
\label{fig:total}
\end{center}
\end{figure}

In graphical form, the imaginary part of the forward $\gamma^{\ast}$-deuteron scattering
amplitude, which is proportional to the deuteron structure function $F_2^D$, is presented in 
Figs.~\ref{fig:ia} and \ref{fig:total}. The dashed vertical lines denote unitary cuts 
placing the 
cut lines and vertices on mass-shell. Figure~\ref{fig:ia} corresponds to the impulse approximation,
when the shadowing correction to $F_2^D$ is neglected. 
Figure~\ref{fig:total} depicts the interference diagram, which gives rise to the shadowing
correction to  $F_2^D$. In this figure, the zigzag lines denote the diffractive scattering
(diffractive exchange).

One can demonstrate that the imaginary part of the interference graph decreases $F_2^D$
(gives the negative shadowing correction) using the Reggeon calculus cutting rules of
Abramovski$\breve{{\rm i}}$, Gribov and Kancheli (AGK), which relate the shadowing effects 
in the total 
and partial cross sections,\cite{AGK} see also Ref.~25.
The top left graph in Fig.~\ref{fig:total} corresponds to the diffractive final state and is
proportional to $(1+\eta^2)({\rm Im} {\cal A})^2 $, where  ${\cal A}$ is the amplitude 
for the photon-nucleon diffractive scattering and $\eta={\rm Re}{\cal A} /{\rm Im}{\cal A}$. 
The top right and bottom left graphs correspond to the inelastic interaction with one of the 
deuteron nucleons and give the commulative contribution proportional to
$-4({\rm Im}{\cal A})^2$. Finally, the bottom right graph corresponds to the 
simultaneous   inelastic
interactions with two  nucleons and is proportional to $2({\rm Im} {\cal A})^2$. Therefore, the sum
of all four graphs in Fig.~\ref{fig:total} is proportional to 
$-(1-\eta^2)({\rm Im} {\cal A})^2=-(1-\eta^2)/(1+\eta^2)F_2^{D(4)}$, where we expressed the
 diffractive amplitude in terms of the
nucleon diffractive structure function $F_2^{D(4)}$.
Therefore, the AGK cutting rules demonstrate that after summing over all final states,
the nuclear shadowing term (the imaginary part of the interference diagram)
 is simply proportional to the contribution of the diffractive final state
 (the top left graph
 in Fig.~\ref{fig:total}).

The deuteron unpolarized 
inclusive structure function $F_2^D$ reads\cite{Frankfurt:2003jf}
\begin{eqnarray}
&&F_2^{D}(x,Q^2)  =  F_{2}^p(x,Q^2)+F_{2}^n(x,Q^2) \nonumber\\
&&-2 \frac{1-\eta^2}{1+\eta^2}\int_{x}^{x_{0}} dx_{\Pomeron}\, d q_t^2
\, F^{D(4)}_2\left(\beta, Q^2,x_{\Pomeron},t\right) \rho_D\left(4 q_t^2+4 (x_{\Pomeron} m_N)^2\right) \,,
\label{sh1}
\end{eqnarray}
where the first line corresponds to the impulse approximation (Fig.~\ref{fig:ia}) and the second
 line corresponds to the interference term (Fig.~\ref{fig:total}).
In this equation, $F_2^{D(4)}$ is  the nucleon diffractive structure function; 
$x_{\Pomeron}$ is the fractional loss of the proton longitudinal momentum;
$\beta \approx x/x_{\Pomeron}$; $x_{0}=0.1$;
 $q=q_t+(x_{\Pomeron} m_N) e_z$ is the momentum transferred to the proton; 
 $\rho_D$ is
 the deuteron charge form factor; $|t|=q_t^2+(x_{\Pomeron} m_N)^2$.
The deuteron charge form factor can be written as an overlap of
the initial and final state deuteron wave functions
\begin{eqnarray}
\rho_D\left(4 q_t^2+4 (x_{\Pomeron} m_N)^2\right)&=&\int d^3 p \left[u(p)u(p+q) \nonumber \right. \\
&&\left. + w(p)w(p+q) \left(\frac{3}{2}\frac{(\vec{p}\cdot (\vec{p}+\vec{q}))}{p^2 (p+q)^2}-\frac{1}{2}\right) \right] \,,
\end{eqnarray}
where $u$ is the $S$-wave component of the deuteron wave function;
$w$ is the $D$-wave component.
Note the argument of the  deuteron form factor, which is a consequence of the correct
treatment of the deuteron center of mass.
Since the $t$-dependence of $\rho_D$ is rather moderate 
(compared to heavier nuclei), the integral in Eq.~(\ref{sh1}) is
sensitive to $F^{D(4)}_2(t)$ up to $-t \leq 0.05$ GeV$^2$.
 The ratio of the real to imaginary parts of the diffractive scattering 
amplitudes $\eta$ can be
calculated using dispersion relations over the  energy or  
using the Regge-pole type parameterization $s^{\alpha_{\Pomeron}}$ 
for the energy 
dependence of the diffraction  cross section,\cite{Migdal}
even though the energy dependence differs from that for soft QCD
processes, 
\begin{equation}
\eta = -\frac{\pi}{2} \frac{\partial \ln(\sqrt{f^D_{i/N}})}{ \partial
 \ln(1/x_{"\Pomeron"})} \approx \frac{\pi}{2} (\alpha_{\Pomeron}(t=0)-1) 
\,,
\label{eta}
\end{equation}
where  $\alpha_{\Pomeron}(0)$ is the intercept of the effective "Pomeron"
trajectory, which differs from that of the actual Pomeron,
which dominates soft QCD phenomena.
Using $\eta \approx 0.32$,\cite{H1:1994} one readily observes that the correction for the real
part of the diffractive scattering amplitude reduces nuclear shadowing by almost 20\%.
Note that in the Reggeon calculus derivation,\cite{Gribov} it was assumed that $\eta=0$,
which is natural for the amplitudes slowly increasing with energy.
This is not the case for DIS and, hence, the effect of $\eta$ should be taken into account.
One should note that the simple final expression for $F_2^{D}$ in Eq.~(\ref{sh1}) is
due to the used closure relation for the final nuclear states.

The use of the QCD factorization theorem for 
hard diffraction\cite{Collins:1997sr} allows to extend Eq.~(\ref{sh1}) for the structure
function $F_2^D$ to
the deuteron parton distribution functions\cite{Frankfurt:1998ym}~$f_{j/D}$
\begin{eqnarray}
&&f_{j/D}(x,Q^2) =  f_{j/p}(x,Q^2)+f_{j/n}(x,Q^2) \nonumber\\
&&-2 \frac{1-\eta^2}{1+\eta^2}
\int_{x}^{x_{0}} dx_{\Pomeron} \,d  q_t^2 \, f^{D}_{j/N}\left(\beta, Q^2,x_{\Pomeron},t\right) \rho_D(4 q_t^2+4 (x_{\Pomeron} m_N)^2) \,.
\label{shdeu}
\end{eqnarray}
Note that we use $x_0=0.1$ for quarks and $x_0=0.03$ for gluons, see the discussion in Ref.~18.

\begin{figure}[ht]
\begin{center}
\epsfig{file=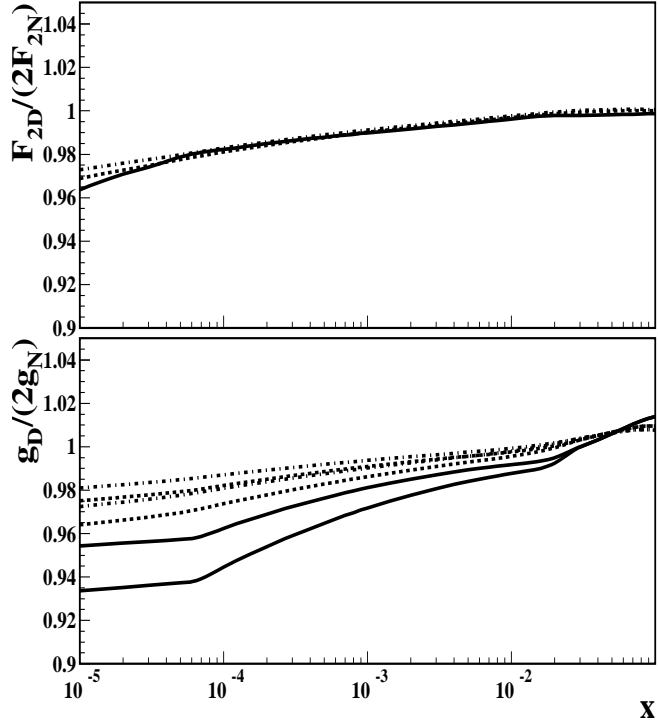,width=9cm,height=11cm}
\vskip -0.3cm
\caption{The ratios $F_2^D/(F_2^p+F_2^n)$ and $g_D/(2g_N)$ as functions of $x$. The solid curve corresponds to $Q=2$ GeV; the dashed curve corresponds to $Q=5$ GeV; the  dash-dotted curve corresponds to $Q=10$ GeV.
Two sets of curves for $g_D/(2g_N)$ reflect the uncertainty in the slope
of the $t$-dependence of the gluon diffractive PDF (see details in the text).}
\label{fig:f2d}
\end{center}
\end{figure}

The results of the calculation of the ratio of the next-to-leading order (NLO) structure
functions $F_2^D/(F_2^p+F_2^n)$ and the ratio of the
NLO gluon PDFs $g_D/(2g_N)$ are presented in Fig.~\ref{fig:f2d}.
The solid curves correspond to $Q=2$ GeV; the dashed curves correspond to $Q=5$ GeV;
 the  dash-dotted curves correspond to $Q=10$ GeV. The two sets of curves for $g_D/(2g_N)$
correspond to the two scenarios of nuclear shadowing for gluons,
see the discussion below and Ref.~19
 for the 
detailed discussion.

As an input for our calculation, we used the H1 fit to the nucleon diffractive
PDFs $f^{D}_{j/N}$,\cite{H1:1994} where the gluon distribution was reduced by the $0.75$ 
factor in accordance with the most recent H1 data on hard inclusive diffraction at HERA.\cite{H1:1997}
Even with this reduction, the gluon nucleon diffractive PDF is much larger than the quark
diffractive PDFs. As a consequence, see Eq.~(\ref{shdeu}), we predict that shadowing in the
gluon channel is larger than that in the quark channel and for $F_2^D/(F_2^p+F_2^n)$.

The $t$-dependence of the quark diffractive PDFs $f^{D}_{j/N}$ was chosen to be
 exponential with the slope\cite{ZEUS:tslope} $B=7.2 \pm 1.1$ GeV$^{-2}$.
Since the $t$-dependence of the gluon PDF has not been measured, 
the slope of the gluon PDF could be different.
 In order to take into account the uncertainty in the value of 
the slope $B$ of the gluon diffractive PDF, we
use two values: $B_g=4+0.2 \ln(10^{-3}/x)$ GeV$^{-2}$ and
$B_g=6+0.25 \ln(10^{-3}/x)$ GeV$^{-2}$. 
The first one corresponds to the lower end of the values of the $J/\psi$ 
photoproduction slope
reported at HERA,\cite{ph1} while the second one is close to the quark slope $B$ and 
to the $J/ \psi$ slope reported in Ref.~30.

The used deuteron wave function corresponds to the Paris nucleon-nucleon potential.\cite{Lacombe:1980dr}

Note that the results in Fig.~\ref{fig:f2d} correspond to the leading twist component of 
nuclear shadowing because they are based on the leading twist analysis of diffraction at HERA.
When one decreases $Q^2$ below $Q^2 \geq 4$ GeV$^2$, for instance down to $Q^2 \leq 1$ GeV$^2$,
one expects a significant enhancement of the nuclear shadowing effect due to the
enhancement of diffraction at  small $Q^2$ by higher twist effects such as vector meson
production. This will increase nuclear shadowing by approximately a factor of 
two.\cite{Frankfurt:2003zd}

Since the diffractive structure function $F_2^{D(4)}$ is known with 
accuracy of approximately 20\%,\cite{H1:1994} 
 the accuracy of the calculation of the nuclear shadowing
 correction to the deuteron structure function $F_2^D$ is $20 \times 0.03=0.6$\%.
 Correspondingly, the theoretical uncertainty
for the ratio of $F_2^n/F_2^p$ extracted from the deuteron data will be 
$2 \times 0.6=1.2$\%, which is likely
to be smaller than possible experimental
systematic errors.

For completeness, we also list predictions for the polarized deuteron structure function $g_1^D$.
Unlike the unpolarized case considered above when the shadowing correction was important for the 
extraction of $F_2^p/F_2^n$ from deuterium data because $F_2^p$ and $F_2^n$ are very close at small $x$,
the polarized $g_1^p \approx -g_1^n$ at small $x$, which makes the shadowing effect a very small 
correction in the extraction of $g_1^n$ from polarized deuteron data, see 
e.g.~Refs.~32 and 33.
In almost complete analogy with the unpolarized case, the deuteron structure function $g_1^D$ can be written
as a sum of the impulse approximation and nuclear shadowing (interference) contributions
\begin{eqnarray}
&&g_1^{D}(x,Q^2)  =  \left(1-\frac{3}{2} P_D \right) \left(g_{1}^p(x,Q^2)+g_{1}^n(x,Q^2) \right) \nonumber\\
&&-2 \frac{1-\eta^2}{1+\eta^2}\int_{x}^{x_{0}} dx_{\Pomeron}\, d q_t^2
\, \Delta F^{D(4)}\left(\beta, Q^2,x_{\Pomeron},t\right) \rho_D^{11} \left(4 q_t^2+4 (x_{\Pomeron} m_N)^2\right) \,,
\label{g1d}
\end{eqnarray}
where $1-3/2 \,P_D$ is the effective polarization of the proton and neutron in the deuteron, which
differs from unity due to the deuteron $D$-wave contribution ($P_D=0.06$ for the Paris nucleon-nucleon potential);
$\rho_D^{11}$ is the electric form factor of the deuteron polarized in the longitudinal direction,
which contains the charge and quadrupole form factor contributions\cite{Frankfurt:1994kt};
$\Delta F^{D(4)}=F^{D(4)}_{\uparrow \uparrow}-F^{D(4)}_{\uparrow \downarrow}$ is the difference
of the diffractive polarized nucleon structure functions. The first arrow stands for the helicity 
of the photon; the second arrow indicates the helicity of the nucleon.
The $\rho_D^{11}$ form factor has the following representation in terms of the deuteron
$S$ and $D$-wave components\cite{Frankfurt:1994kt}
\begin{eqnarray}
&&\rho_D^{11}\left(4 q_t^2+4 (x_{\Pomeron} m_N)^2\right)=\int d^3 p \left[u(p)u(p+q) \nonumber \right. \\
&& \left. + \frac{u(p)w(p+q)}{\sqrt{2}}\left(\frac{3}{2}\frac{(p_z+q_z)^2}{(p+q)^2}-\frac{1}{2}\right)+
\frac{u(p+q)w(p)}{\sqrt{2}}\left(\frac{3}{2}\frac{p_z^2}{p^2}-\frac{1}{2}\right) \nonumber \right. \\
&& \left. + w(p)w(p+q) \left(\frac{9}{2} \frac{(\vec{p}_t \cdot(\vec{p}_t+\vec{q}_t)) (\vec{p} \cdot
(\vec{p}+\vec{q}))}{p^2 (p+q)^2} +\frac{3}{4}\frac{p_z^2}{p^2}+\frac{3}{4}\frac{(p_z+q_z)^2}{(p+q)^2}-1
\right) \right] \,.
\end{eqnarray}

Since $\Delta F^{D(4)}$ is a new and unmeasured quantity 
(it can be measured in polarized diffractive DIS on the nucleon), 
we cannot directly use the leading twist theory of nuclear shadowing 
to estimate the shadowing correction to $g_1^{D}$. However, 
{\it making an assumption} that the relative
strength of diffraction mediated by the non-vacuum
exchange (responsible for the polarized structure function $g_1$ at small $x$) is the same
as that of the exchange with vacuum quantum numbers (responsible for the unpolarized
$F_2$ at small $x$), one obtains that
\begin{equation}
\frac{\Delta F^{D(4)}}{g_1^N}=2 \frac{F_2^{D(4)}}{F_2^N} \,,
\label{eq:assumption}
\end{equation}
where $g_1^N=(g_1^p+g_1^n)/2$ and $F_2^N=(F_2^p+F_2^n)/2$.
This assumption allows one to express the shadowing correction to
the ratio of the deuteron and nucleon spin structure functions in terms of unpolarized diffraction on the nucleon
\begin{eqnarray}
\frac{g_1^{D}(x,Q^2)}{2 \left(1-\frac{3}{2} P_D \right) g_1^N}&&=1-4 \frac{1-\eta^2}{1+\eta^2} 
\frac{1}{\left(1-\frac{3}{2} P_D \right)}\frac{1}{F_2^N(x,Q^2)} \nonumber\\
&&\times \int_{x}^{x_{0}} dx_{\Pomeron}\, d q_t^2
\, F_2^{D(4)}\left(\beta, Q^2,x_{\Pomeron},t\right) \rho_D^{11} \left(4 q_t^2+4 (x_{\Pomeron} m_N)^2\right) \,.
\label{g1d:ratio}
\end{eqnarray}
The assumption of Eq.~(\ref{eq:assumption}) corresponds to the maximal shadowing correction. 
The additional factor of two is a source of the generic combinatoric enhancement of nuclear shadowing in
polarized structure functions of few-nucleon nuclei compared to the unpolarized case.
\cite{Frankfurt:1996nf}

\begin{figure}[h]
\begin{center}
\epsfig{file=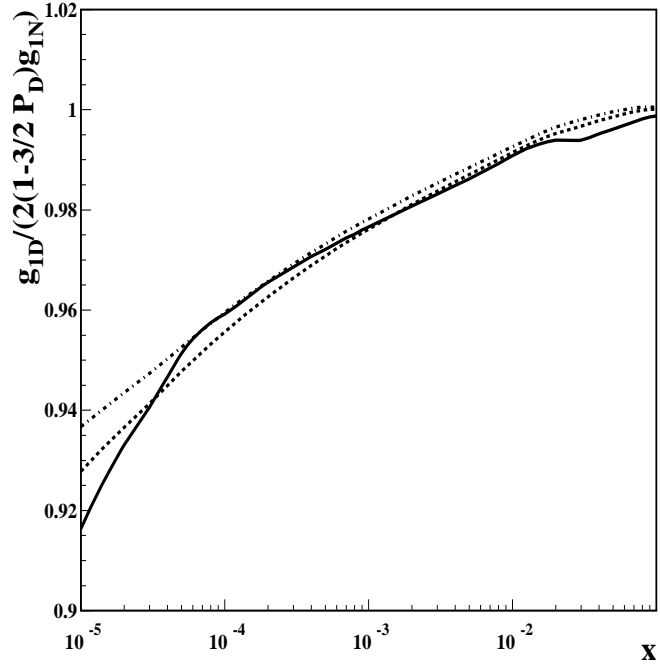,width=9cm,height=10cm}
\vskip -0.3cm
\caption{The ratio $g_1^D/[(1-3/2\, P_D)g_1^N]$ as a function of $x$. The solid curve corresponds to $Q=2$ GeV; the dashed curve corresponds to $Q=5$ GeV; the  dash-dotted curve corresponds to $Q=10$ GeV.}
\label{fig:g1d}
\end{center}
\end{figure}

Figure~\ref{fig:g1d} presents the results of the calculation using Eq.~(\ref{g1d:ratio}). The
solid curve corresponds to $Q=2$ GeV, the dashed curve corresponds to $Q=5$ GeV and
the dot-dashed curve corresponds to $Q=10$ GeV.

As can be seen from Fig.~\ref{fig:g1d}, the shadowing correction to 
$g_1^D$ could be as large as 8\% at $x \approx 10^{-5}$. 
Needless to say that in order to achieve such low values of Bjorken $x$ 
simultaneously with $Q^2 \geq 1$ GeV$^2$, one needs a collider with the 
polarized deuteron beam. The available fixed-target data\cite{Ageev:2005gh} 
can probe $g_1^D$ only down to $x \approx 0.004$, where the shadowing correction
is very small.

It was noticed in Ref.~16
 that for a spin-one target
(deuteron), the cross section of DIS depends on the deuteron polarization
even with the unpolarized beam. The associated asymmetry
\begin{equation}
T_{20}=\frac{\sigma^{+}-\sigma^{0}}{\frac{1}{2}(\sigma^{+}+\sigma^{0})} \,,
\label{eq:t20}
\end{equation}
where $\sigma^{+,0}$ denotes the $\gamma^{\ast}$-deuteron cross section
and the superscript denotes the deuteron helicity, was estimated
for $x > 0.1$ in the impulse approximation.\cite{Frankfurt:1981mk} 
Next, it was pointed out in Ref.~37
 that nuclear shadowing 
in unpolarized DIS on deuterium leads to the values of the $T_{20}$ asymmetry at the 
level of one percent at small-$x$.

This can be estimated as follows. The definition~(\ref{eq:t20}) 
allows one to
immediately write the expression for $T_{20}$ by replacing the deuteron charge form factor
in Eq.~(\ref{sh1})  by $\rho_D^{20}$, 
\begin{eqnarray}
&&T_{20}(x,Q^2)= \nonumber\\
&&\frac{2}{F_2^D(x,Q^2)}\frac{1-\eta^2}{1+\eta^2}\int_{x}^{x_{0}} dx_{\Pomeron}\, d q_t^2
\, F^{D(4)}_2\left(\beta, Q^2,x_{\Pomeron},t\right) \rho_D^{20}\left(4 q_t^2+4 (x_{\Pomeron} m_N)^2\right) 
\label{eq:t20_b}
\end{eqnarray}
where 
\begin{eqnarray}
&&\rho_D^{20}\left(4 q_t^2+4 (x_{\Pomeron} m_N)^2\right)=\frac{3}{2} \int d^3 p 
\Big[\frac{u(p)w(p+q)}{\sqrt{2}} \left(1-\frac{3(p_z+q_z)^2}{(p+q)^2}\right) \nonumber \\ 
&& +\frac{u(p+q)w(p)}{\sqrt{2}} \left(1-\frac{3\,p_z^2}{p^2}\right) 
+w(p)w(p+q)\Big(1 -\frac{3}{2} \Big[\frac{(p_z+q_z)^2}{(p+q)^2}+\frac{p_z^2}{p^2} \nonumber \\ 
&& +\frac{(\vec{p} \cdot (\vec{p}+\vec{q})) (\vec{p} \cdot (\vec{p}+\vec{q})-3 p_z (p_z+q_z))}{p^2 (p+q)^2}\Big]\Big)\Big] \,.
\end{eqnarray}
Note that $T_{20}$ vanishes, if one ignores the $D$-wave component of the
 deuteron wave function or the nuclear shadowing correction.

This effect can be also formulated in terms of the third deuteron
structure function, $b_1^D$, which has the following probabilistic interpretation in terms
of the quark distributions\cite{Hoodbhoy:1988am}
\begin{equation}
b_1^D(x)=\frac{1}{2} \sum e_q^2 \left[q^0(x)+\bar{q}^0(x)-\frac{1}{2}\left(q^1(x)+\bar{q}^1(x)+q^{-1}(x)+\bar{q}^{-1}(x)\right)\right] \,,
\label{eq:b1}
\end{equation}
where $q^{\lambda}$ is the unpolarized quark distribution in the deuteron with helicity $\lambda$. The connection between $b_1^D$ and $T_{20}$ is readily obtained using their definitions,
\begin{equation} 
b_1^D(x,Q^2)=-\frac{F^D_2(x,Q^2)}{2\, x} T_{20}(x,Q^2) \,.
\label{eq:b1t20}
\end{equation}
The factor $1/(2x)$ in Eq.~(\ref{eq:b1t20}) indicates that the often discussed $b_1^D$
structure function is a rather inappropriate quantity: even small values of the physically
measured $T_{20}$ asymmetry correspond to huge values of $b_1^D$.

\begin{figure}[h]
\begin{center}
\epsfig{file=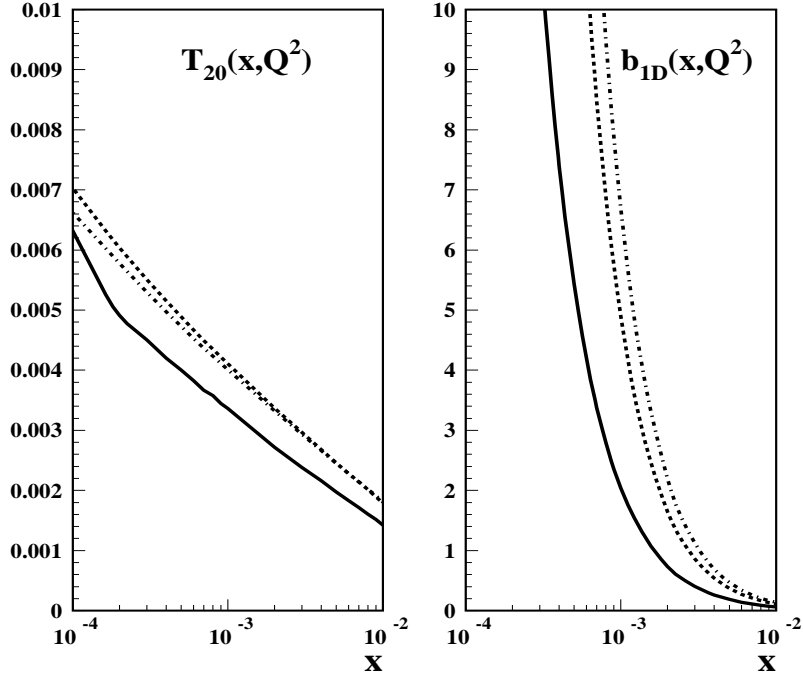,width=11cm,height=10cm}
\vskip -0.0cm
\caption{The tensor asymmetry $T_{20}$ and the $b_1^D(x,Q^2)$ structure function as functions of $x$. The solid 
curve corresponds to $Q=2$ GeV; the overlapping dashed and dash-dotted curves correspond to $Q=5$ GeV and $Q=10$ GeV.}
\label{fig:b1}
\end{center}
\end{figure}

The results of the calculation of the tensor asymmetry $T_{20}$ and
the deuteron structure function $b_1^D$ are presented in
Fig.~\ref{fig:b1}. The solid curve corresponds to $Q=2$ GeV; the overlapping dashed and dash-dotted
 curves correspond to $Q=5$ GeV and $Q=10$ GeV.

As one can see from the left panel of Fig.~\ref{fig:b1}, the obtained  
$T_{20}$ is at the level of 0.6-0.7\%. This agrees with the 
the analyses of Refs.~32 and 37.
At the same time, $b_1^D$ is large at small $x$, as can be seen from the right panel of
Fig.~\ref{fig:b1}. This is a purely kinematic effect
due to the $1/(2\,x)$ factor in the definition of $b_1^D$~(\ref{eq:b1t20}).
The observation of surprisingly large $b_1^D$ at small $x$ was first presented 
in Refs.~32, 33 and 39.

The HERMES measurement\cite{Airapetian:2005cb} of  $b_1^D$ indicates a (rapid) 
growth of $b_1^D$ when one
decreases Bjorken $x$ from $x \approx 0.5$ down to $x \approx 10^{-2}$.
However, the corresponding values of $Q^2$ are of the order of 1 GeV$^2$ and
the values of Bjorken $x$ are not small enough to see the predicted dramatic rise of 
$b_1^D$ towards small $x$. Once again, the study of the  behavior of  $b_1^D$ at small $x$
will greatly benefit from the collider kinematics.

\section{Nuclear shadowing and final state interactions in the tagged deuteron structure
function}
\label{sec:tagged}

A   strategy, which is complimentary to the inclusive measurement of $F_2^{D}$,
is the use of the neutron and proton tagging. The scattering on the neutron of
deuterium is then tagged by detecting a slow (spectator) proton.
The usefulness of the tagged deuteron structure function for the extraction of the neutron 
$F_2^n$ at large $x$ was discussed in Ref.~41.
 In this work, we concentrate
on the small-$x$ region of nuclear shadowing. We extend the analysis\cite{Frankfurt:2003jf}
 by taking into account the
final state interactions (FSI) between the final nucleons.

\begin{figure}[h]
\begin{center}
\epsfig{file=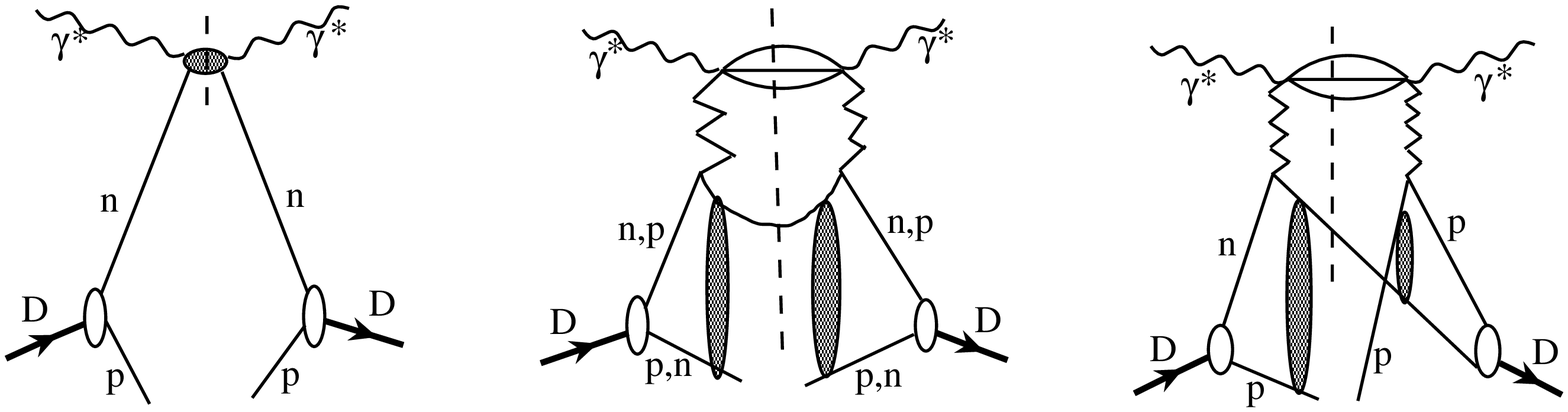,width=13cm,height=6cm}
\vskip 0cm
\caption{The impulse approximation and the final state correction to the tagged
deuteron structure function $F_2^D$.}
\label{fig:tagged_impulse}
\end{center}
\end{figure}

In the impulse approximation, the tagged deuteron structure function is given by the imaginary part
of the left graph in Fig.~\ref{fig:tagged_impulse}.

The corresponding expression is
\begin{equation}
F_2^D(x,Q^2,\vec{p})\Big|_{{\rm IA}}=\left(1+\frac{p_z}{m_N}\right)^{\alpha_{\Pomeron}(0)}\, F_2^n(x,Q^2)\, \rho_D(p,p) \,,
\label{ia}
\end{equation}
where $\vec{p}$ is the momentum of the spectator proton; $\alpha_{\Pomeron}(0)$ is the
intercept of the effective ``Pomeron'' trajectory;
$\rho_D(p,p)=u^2(p)+w^2(p)$, where
$u(p)$ and $w(p)$ are the $S$-wave and the $D$-wave components of the deuteron wave function,
is the unpolarized deuteron density matrix\cite{Frankfurt:1994kt} (with equal momenta).
The factor $(1+p_z/m_N)^{\alpha_{\Pomeron}(0)}$ comes from different invariant energies
of the virtual photon-nucleus and the virtual photon-neutron interactions. 
Thus, the $(1+p_z/m_N)^{\alpha_{\Pomeron}(0)}$ factor is the flux factor of the interacting neutron.
Note that in the derivation of the $(1+p_z/m_N)^{\alpha_{\Pomeron}(0)}$
 factor we neglected ${\cal O}(p^2/m_N^2)$ and
higher corrections, which are ignored in our non-relativistic treatment of the
deuteron wave function.

The impulse approximation receives corrections due to the final state interactions 
(the middle graphs in Fig.~\ref{fig:tagged_impulse}) and
nuclear shadowing, which is also accompanied by the final state interactions (the right graph
in Fig.~\ref{fig:tagged_impulse}). The shaded ovals in Fig.~\ref{fig:tagged_impulse} denote the
final state interaction. Since the FSI between two nucleons is largest 
at the small relative momentum, the FSI is accompanied by the diffractive
scattering providing this condition.

In the case of the tagged structure function at small spectator nucleon momenta,
 $p \le \sqrt{\epsilon m_N}$, where $\epsilon$ is the deuteron binding energy, 
it is legitimate to keep only the single and double scattering terms shown in
Fig.~\ref{fig:tagged_impulse}. At larger spectator momenta, 
the additional contributions of triple and quadruple interactions with the target
(not shown in Fig.~\ref{fig:tagged_impulse}) are not suppressed by
 the $p / \sqrt{\epsilon m_N}$ parameter and, hence, should be included.
 This will introduce a certain model dependence since those terms
are not simply related to the elementary diffractive cross section.
One should emphasize that this is only the case for the tagged structure function
(differential cross section):
the triple and quadruple interaction terms cancel in the inclusive structure
function (total cross section), which is unambiguously expressed in terms of the
nucleon diffractive structure function, see Eq.~(\ref{sh1}).

The complete expression for the tagged deuteron structure function reads
\begin{eqnarray}
&&F_2^D(x,Q^2,\vec{p})=\left(1+\frac{p_z}{m_N}\right)^{\alpha_{\Pomeron(0)}}\, F_2^n(x,Q^2)\, \rho_D(p,p) \nonumber \\
&+&\int_{x}^{x_{0}} dx_{\Pomeron}\, d q_t^2 \, F_2^{D(4)}\left(\beta, Q^2,x_{\Pomeron},t\right)
\Big[\Big( 2 Re \int d^3 p^{\prime}\Big({\rm F}(p,p^{\prime})\,\rho_D(p,p^{\prime})\nonumber\\
&+&{\rm F}(p+q,p^{\prime})\,\rho_D(p+q,p^{\prime}) \Big) \nonumber\\
&+&2 \int d^3 p^{\prime} \int d^3 p^{\prime \prime}{\rm F}(p^{\prime},p^{\prime \prime})\, \rho_D(p^{\prime},p^{\prime \prime})\psi_{NN}^{{\rm FSI}}(p-q/2;p^{\prime}-q/2)\Big) \psi_{NN}^{{\rm FSI}}(p-\frac{q}{2};p^{\prime \prime}-\frac{q}{2})
\Big] \nonumber\\
&-&2 \frac{1-\eta^2}{1+\eta^2} \int_{x}^{x_{0}} dx_{\Pomeron}\, d q_t^2 \, F_2^{D(4)}\left(\beta, Q^2,x_{\Pomeron},t\right)
\Big({\rm F}(p,p+q)\,\rho_D(p,p+q) \nonumber\\
&+&Re\int d^3 p^{\prime} \left({\rm F}(p,p^{\prime})\,\rho_D(p,p^{\prime})+{\rm F}(p+q,p^{\prime})\,\rho_D(p+q,p^{\prime})\right)
 \psi_{NN}^{{\rm FSI}}(p-\frac{q}{2};p^{\prime}-\frac{q}{2}) \nonumber\\
&+& \int d^3 p^{\prime} \int d^3 p^{\prime \prime}{\rm F}(p^{\prime},p^{\prime \prime})\, \rho_D(p^{\prime},p^{\prime \prime})\psi_{NN}^{{\rm FSI}}(p-\frac{q}{2};p^{\prime}-\frac{q}{2}) \psi_{NN}^{{\rm FSI}}(p-\frac{q}{2};p^{\prime \prime}-\frac{q}{2}) \Big) \,,
\label{spec:full}
\end{eqnarray}
where $\psi_{NN}^{{\rm FSI}}$ is the continuum non-relativistic nucleon-nucleon wave function, which vanishes
 in the absence of
the FSI. The arguments of $\psi_{NN}^{{\rm FSI}}$ denote the final and initial relative momenta of the involved proton
and neutron. 
The $\rho_D(p,p^{\prime})$ denotes the deuteron unpolarized density matrix
\begin{equation}
\rho_D(p,p^{\prime})=u(p)u(p^{\prime})+w(p)w(p^{\prime})\left(\frac{3}{2} \frac{(\vec{p} \cdot \vec{p^{\prime}})^2}{p^2 p^{\prime 2}}-\frac{1}{2} \right)  \,.
\end{equation}
The factor ${\rm F}(p,p^{\prime})$ is the generalization of the nucleon flux factor
discussed above
\begin{equation}
{\rm F}(p,p^{\prime})=\sqrt{\left(1+\frac{p_z}{m_N}\right)^{\alpha_{\Pomeron(0)}}
\left(1+\frac{p^{\prime}_z}{m_N}\right)^{\alpha_{\Pomeron(0)}}} \,.
\end{equation}
Note that the presence of the flux factor is typical for semi-inclusive cross sections.
In total cross sections, the flux factor effects cancel in the impulse approximation and are
of the order ${\cal O}(p^2/m_N^2)$ in the interference term. Therefore, they do not appear in 
Eqs.~(\ref{sh1}), (\ref{g1d}) and (\ref{eq:b1}).

By setting $\psi_{NN}^{{\rm FSI}}=0$ in Eq.~(\ref{spec:full}), we obtain the result of of Ref.~15.
The second and third lines in Eq.~(\ref{spec:full}) correspond to the FSI correction
to the impulse approximation (the middle graph of Fig.~\ref{fig:tagged_impulse}; the rest of
 Eq.~(\ref{spec:full}) corresponds to the shadowing correction, which also includes the FSI.

In our numerical analysis of Eq.~(\ref{spec:full}),
we make the following justified approximations. First, since the 
dominant diffractive exchange at small $x$ 
has the vacuum quantum numbers, the isospin of the interacting proton-neutron pair
is the same as in the deuteron, i.e.~it is zero.
Note that the approximation of the exchange with the vacuum quantum numbers is
only justified for small $x$, $x < 10^{-2}$. At larger $x$, one should also
take into account transitions with the isospin change, which makes an estimate
of the final state interactions more involved.
 Second, an examination of the 
 isospin-zero proton-neutron phase shifts\cite{Arndt:2000xc,Machleidt:2000ge}
reveals that, in the appropriate kinematics (small kinetic 
energy in the laboratory frame), the only essential partial wave is
the $^{3}S_1$-wave, i.e.~the partial wave with $L=0$ and $S=1$.
Therefore, this is the only partial wave included 
in our analysis of the FSI.
This allows us to suppress all isospin and spin indices in Eq.~(\ref{spec:full}).

The $L=0$ relative orbital momentum allows for a simple expression of the 
nucleon-nucleon continuum wave function in the momentum space
\begin{equation}
\psi_{NN}(k_1,k_2)=\delta^3(\vec{k}_1-\vec{k}_2)+\psi_{NN}^{{\rm FSI}}(k_1,k_2)
=\delta^3(\vec{k}_1-\vec{k}_2)+\frac{(e^{2i \delta_{01}}-1)}{4 \pi^2 i |k_1|} \,, 
\label{eq:wf_general}
\end{equation}
where $\delta_{01}$ is the phase shift of the  $^{3}S_1$-wave partial wave.
The wave function of Eq.~(\ref{eq:wf_general}) has the usual normalization of
continuum wave functions in non-relativistic quantum mechanics
\begin{equation}
\int d^3 \vec{k}_2 \,\psi_{NN}^{\ast}(k_1^{\prime},k_2)\, \psi_{NN}(k_1,k_2)=\delta^3(\vec{k}^{\prime}_1-\vec{k}_1) \,.
\label{eq:normalization}
\end{equation}
Since $\delta_{01}$ is close to $90^0$ in the considered 
kinematics,\cite{Arndt:2000xc,Machleidt:2000ge} 
$\psi_{NN}^{{\rm FSI}}$ reduces to the following simple expression, which we used in
our numerical analysis
\begin{equation}
\psi_{NN}^{{\rm FSI}}(k_1,k_2) \approx \frac{\left(\cos \,(2 \,\delta_{01})-1 \right)}{4 \pi |k_1|} \delta(|k_1|^2-|k_2|^2)  \,.
\label{eq:swave}
\end{equation}
Note that the continuum wave function of Eqs.~(\ref{eq:wf_general}) and (\ref{eq:swave}) with $\delta_{01} \approx 90^0$ 
agrees with the classic Bethe-Pierls result for the low relative momentum $k$  proton-neutron
continuum wave function in the coordinate space
\begin{equation}
\psi_{NN}^{{\rm FSI}}(r) \propto \frac{\sin(kr +\delta_0)}{r} \,,
\label{eq:bethe}
\end{equation}
where $\cot \delta_0=-\sqrt{\epsilon m_N}/k$ with $\epsilon=2.15$ MeV being the deuteron
binding energy.\cite{Bethe} Assuming that the relevant momentum $k$ equals 
the root-mean-square momentum
of the deuteron wave function, $k \approx 130$ MeV for the used Paris nucleon-nucleon potential,
we find that the phase shift $\delta_0 \approx 110^0$, which is in agreement with the value used our analysis.

In our analysis, we used 
the parameterization of $\delta_{01}$ provided by the SAID program.\cite{SAID}
Other applications of the non-relativistic nucleon-nucleon wave function to estimates of
the FSI can be found in Refs.~47 and 48.

The coefficient in front of the nuclear shadowing correction term, $-2\,(1-\eta^2)/(1+\eta^2)$,
is dictated by the AGK cutting rules. Indeed, 
in the considered case of the proton tagging,
 the shadowing correction is given by the sum of the two top graphs  in Fig.~\ref{fig:total}, 
which enter with the coefficients $2$ and $-4/(1+\eta^2)$, respectively.

\begin{figure}[h]
\begin{center}
\epsfig{file=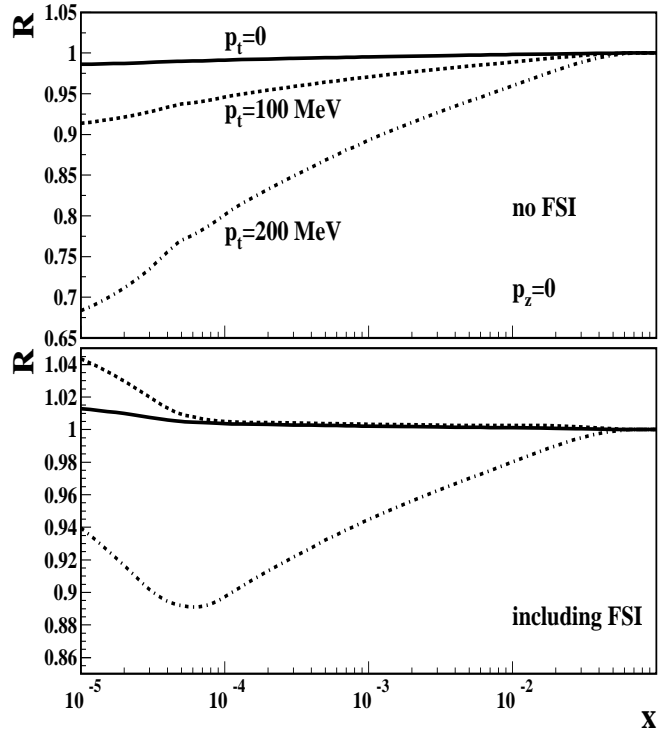,width=9cm,height=11cm}
\vskip 0.0cm
\caption{The ratio $R$ of Eq.~(\ref{spec:full}) giving the suppression of the proton spectrum
 by the FSI and nuclear shadowing. The top panel corresponds to the calculation 
without the FSI ($\psi_NN=0$); the lower panel is the result of the full calculation.
 The solid curves corresponds to $p_t=0$; the  short dashed curves correspond to $p_t=100$ MeV;
 the dash-dotted curves correspond to  $p_t=200$ MeV.}
\label{fig:spec}
\end{center}
\end{figure}

Equation~(\ref{spec:full}) describes the modification of the spectrum of the produced protons
by the FSI and nuclear shadowing. We characterize the modification by the
ratio $R$ defined as
\begin{equation}
R(x,Q^2,\vec{p})=\frac{F_2^D(x,Q^2,\vec{p})}{F_2^D(x,Q^2,\vec{p})\Big|_{{\rm IA}}} \,.
\end{equation}
The results of the numerical evaluation of the ratio $R$ at fixed $Q^2=4$ GeV$^2$ and fixed 
$p_t$ and $p_z=0$ as a function of $x$ are presented in Fig.~\ref{fig:spec}.
 The top panel corresponds to the calculation 
without the FSI [$\psi_{NN}^{{\rm FSI}}=0$ in Eq.~(\ref{spec:full})]; the lower panel is the result of the full calculation.
 The solid curves corresponds to $p_t=0$; the  short dashed curves correspond to $p_t=100$ MeV;
 the dash-dotted curves correspond to  $p_t=200$ MeV.

As one can see from the top panel of Fig.~\ref{fig:spec}, 
nuclear shadowing (without the FSI effects) suppresses the spectrum of the produced protons.
The suppression is larger for larger $p_t$ and becomes as large as 20-30\% at $p_t=200$ MeV and
$x=10^{-5}-10^{-4}$. This is much larger than the 
shadowing correction to the inclusive deuteron structure function $F_2^D$.
This enhancement of nuclear shadowing is a common feature of semi-exclusive reactions with 
nuclei. Indeed, at large $p_t$, while the impulse approximation term is suppressed by
 the nuclear wave function, the rescattering term becomes increasingly prominent.

As can be seen from the bottom panel of Fig.~\ref{fig:spec}, the large effect of nuclear shadowing
 on $R$ is mostly compensated by the final state interactions. A numerical analysis of 
 Eq.~(\ref{spec:full}) shows that the main contribution to the FSI effect comes from the terms
proportional to $\psi_{NN}^{{\rm FSI}}$ and that the terms containing $(\psi_{NN}^{{\rm FSI}})^2$ can be neglected. 
The net result is the interplay between the large and negative contributions of 
$\psi_{NN}^{{\rm FSI}}$ to $R$, which nearly cancel each other because of  the opposite signs
 of the FSI correction to the impulse approximation and the nuclear shadowing term, 
and the shadowing term without the FSI.

From the experimental point of view, two strategies of the extraction of 
the neutron $F_2^n$ from the deuteron data by using the proton tagging 
will be possible. One would be to select only very low $p_t$ protons with a gross loss of statistics.
According to our analysis, the distortion of the proton spectrum due to the
final state interactions and FSI effects will be minimal. The other, more promising
approach  is to measure the $p_t$ dependence of the spectrum up to $p_t \sim 200$ MeV/c, which 
will allow to use most of
the spectator protons. Provided good momentum resolution of the  proton spectrometer, 
one would be able to make (longitudinal) momentum 
cuts to suppress/increase  the shadowing effect and, thus, one
would have an opportunity to independently study the enhanced nuclear shadowing.
The tagged method will allow to extract $F_2^n$ from deuteron data with the accuracy 
at  the level of a fraction of percent.

One can also use simultaneous tagging of protons and neutrons, when 
both neutron and proton  are detected in the reactions
 $\gamma^{\ast} D \to n X$ and $\gamma^{\ast} D \to p X$.
In this case, the effects of
nuclear shadowing and FSI
will cancel in the ratio $\sigma^{\gamma^{\ast} D \to n X}/\sigma^{\gamma^{\ast} D \to p X}$
and the main errors in the measurement of $F_2^n$ will be due to the
determination of relative efficiencies of the proton and  neutron taggers.

One could also try to obtain the ratio $F_2^n /F_2^p$ from the comparison
 of the rate of the tagged proton scattering events with the neutron spectator 
to inclusive $e\,D$ scattering.  Such a strategy could also have certain merits
as it avoids the issue of luminosity and does not require a leading proton spectrometer.
 The disadvantage of this strategy 
is the sensitivity to the nuclear shadowing and FSI effects and errors in the acceptance of
the neutron detector.  One possible way to deal with the latter problem will be to perform 
measurements at very small $x$ and large
energies,  where the
$ep$ and $en$ cross sections are equal 
to better than a fraction of 1\% and, hence, one would be able to cross-check
the  acceptances of the proton and neutron detectors.

Note also that taking proton data from an independent run will 
potentially lead to another
set of issues such as relative luminosity, the use of different
 beam energies, etc., which is likely to be at the level of 1\%.

\section{Conclusions and discussion}
\label{sec:conclusions}

In this brief review, we presented predictions of the theory of leading 
twist nuclear shadowing for inclusive unpolarized and polarized
deuteron structure functions $F_2^D$ and $g_1^D$
and for the tensor polarization asymmetry $T_{20}$.
The combined role of nuclear shadowing and final state interactions was
analyzed for the tagged deuteron structure function $F_2^D(x,Q^2,\vec{p})$.
The discussed effects are relevant for the
collider kinematics of the future Electron-Ion Collider, $x < 10^{-2}$ and
$Q^2  \geq 4$ GeV$^2$.

The measurement of $F_2^D$ is relevant for the extraction of the neutron 
$F_2^n$ structure function from the deuteron data at small $x$. 
Since at small $x$, the proton and neutron $F_2$ structure functions are close,
even a small correction to $F_2^D$ leads to large corrections to the extraction of 
the $F_2^p-F_2^n$ difference for the deuteron data.
We make predictions for the shadowing correction to $F_2^D$, which 
enable one to extract $F_2^p/F_2^n$ from the deuteron data with an approximately 
1\%
theoretical uncertainty.

At moderately small $x$, nuclear shadowing of structure functions of light nuclei 
is dominated by the double rescattering. Thus, the theoretical analysis of $F_2^A/F_2^D$, where 
$F_2^A$ denotes the structure function of a nucleus heavier than deuterium,
 presents another option
for the extraction of the nuclear shadowing correction to  $F_2^D$.\cite{Frankfurt:2003zd} 
This imposes additional constraints on the shadowing correction  to $F_2^D$.

For completeness, we also consider polarized deuteron structure functions.
While the shadowing correction to $g_1^D$ is approximately twice as large as the
shadowing correction to $F_2^D$, its effect is totally negligible in the
extraction of the neutron $g_1^n$ from the deuteron data since 
$g_1^n \approx -g_1^p$ at small $x$ and since the experimental errors in
the measurement of $g_1^D$ are rather large.  
We find that the $T_{20}$ asymmetry  at small-$x$ is at the one percent 
level so that one could try to observe it experimentally.

The strategy complimentary to the inclusive measurement of $F_2^D$ is the measurement
of the proton tagging. We analyzed the deuteron tagged structure function and showed
that the combined effect of nuclear shadowing and the final state 
interactions only insignificantly distorts the spectrum of final protons
for $p_t \leq 200$ MeV, which enables for
a reliable extraction of $F_2^n$.
A combined analysis of inclusive and semi-inclusive scattering off the deuteron 
coupled  with a high resolution proton spectrometer 
will allow for the measurement of $F_2^n$ at small $x$ with the theoretical
uncertainty at the level of 1\%.
 It would be a challenge to reduce
the experimental systematic errors to a comparable level.
The measurement of the shape of the spectator spectrum would allow to
determine nuclear shadowing in 
deuterium with precision by far
exceeding that possible in  the inclusive measurements.

This work was supported by the German-Israel Foundation (GIF), Sofia Kovalevskaya 
Program of the the Alexander
von Humboldt Foundation (Germany)  and the Department of Energy (USA).

\end{document}